\documentclass[prl,twocolumn,aps]{revtex4}
\usepackage{graphicx}
\begin{document}
\title{Thermal conductivity of a classical one dimensional spin-phonon system}
\author{ A.V. Savin$^1$, G. P. Tsironis$^2$ and X. Zotos$^2$}
\affiliation{
$^{1}$ Institute of Chemical Physics RAS, Kosygin str.4, Moscow 119991, 
Russia\\
$^{2}$ Department of Physics, University of Crete and   
Foundation for Research
and Technology-Hellas, P.O. Box 2208, 71003 Heraklion, Greece}
\date{\today}

\begin{abstract}
We investigate the thermal conductivity $\kappa$ 
of the classical Heisenberg spin chain 
coupled to a variety of phonon systems using a Green-Kubo approach
as well as coupling to heat reservoirs. 
The decay of energy current correlations is power law or exponential, 
depending on the type of (an)harmonicity of the phonon system and 
the presence of a substrate. 
In particular, we find that coupling the spin system to 
displacements with an acoustic type harmonic inter-site potential leads 
to a diverging $\kappa$. Adding a substrate, harmonic or 
periodic such as sine-Gordon, drives the system to diffusive behavior.
With anharmonic potential without a substrate the generic transport is 
diffusive even for the integrable Toda potential, while for the FPU 
chain is ballistic.
\end{abstract}

\maketitle

Recent experiments\cite{hess,ott} convincingly 
promoted magnetic excitations as a very efficient mechanism for energy 
transport in quasi-one dimensional materials.  
In particular, spin-1/2 Heisenberg chain and ladder 
(undoped or hole doped) as well as spin-1 compounds have been studied. 
The observed highly anisotropic 
thermal conductivity, that is comparable in magnitude to that of 
metallic systems, was attributed to magnetic transport.
Theoretically, it has been noticed that  
in the one dimensional (1D) spin-1/2 Heisenberg model the energy current 
commutes with the Hamiltonian and thus the thermal conductivity is 
ballistic at all temperatures. 
This observation falls in line with a proposal of unconventional 
transport in 1D integrable systems\cite{review}.

In general, the energy transport in magnetic materials has 
a phonon (lattice) and a magnetic contribution. 
The thermal conductivity is limited by 
phonon-phonon (anharmonicity), spin-spin and spin-phonon scattering. 
In the materials discussed above, described by the 1D spin-1/2 Heisenberg 
Hamiltonian (assuming that there is no disorder), in the absence of spin-spin 
scattering, the thermal conductivity is only limited by phonon-phonon and 
spin-phonon scattering. Clearly there is experimental and theoretical 
interest in understanding the thermal conductivity in this singular 
case\cite{shimshoni,chernyshev,unpublished}. 
However, the generic case of 1D magnetic 
thermal transport for higher or even classical spin (large $S$ limit) is 
a benchmark that is also of experimental interest, e.g. for $S=1,3/2,5/2$  
compounds\cite{ott}.

In this work we foucs on the classical Heisenberg model coupled 
to lattice displacements. As both systems are classical we can resort to large 
scale, state of the art numerical simulations. This is in contrast to the 
quantum problem which is numerically intractable because of the 
unbounded phonon spectrum and thus limited to perturbative or 
semi-phenomenological analysis. 

There is a great deal of work done on the thermal conductivity of 1D 
classical lattices\cite{livi} where a variety of behaviors (ballistic vs. 
diffusive) was established depending on the anharmonicity of the lattice and 
presence of a substrate.
For the classical Heisenberg model it seems that the thermal 
transport is diffusive\cite{muller,stz}.

In the following, we present a series of simulations exploring the role of 
the phonon bath transport properties to the total thermal conductivity.
In particular we find that the coupling to acoustic phonons leads to 
a diverging $\kappa$. This is in contrast to conclusions for 
the $S=1/2$ quantum chain\cite{shimshoni,unpublished} based on the 
memory function approach.
This difference is surprising because we would 
expect that the quantum spin chain (with ballistic transport) would be more 
robust to coupling to phonons than the classical one.

We consider the prototype classical Heisenberg spin chain Hamiltonian 
coupled to lattice degrees of freedom carrying the spins, 
\begin{eqnarray}
H&=&\sum_{n}J_{n,n+1}({\bf S}_n \cdot {\bf S}_{n+1})
+\frac{1}{2}M(d{Q}_n/dt)^2\nonumber\\
&+&Mv_0^2\left\{V([Q_{n+1}-Q_n]/l)+U(Q_n/l)\right\}
\label{h}
\end{eqnarray}
where, ${\bf S}_n=S{\bf e}_n$ (${\bf e}_n$ a unit vector), 
$Q_n$ is the displacement of the $n$th atom with mass $M$ from the equilibrium position
$nl$, $l$ is the lattice spacing, 
$v_0$ is the velocity of longitudinal sound, $V$ ($U$) the 
dimensionless interatomic (substrate) potential ($V(0)=V'(0)=0$, $V''(0)=1$,
$U(0)=U'(0)=0$, $U''(0)=k\ge 0$). 
The exchange constant $J_{n,n+1}$ depends on the distance between the $n$th
and $(n+1)$th atoms as $J_{n,n+1}=J[1-\chi(Q_{n+1}-Q_n)/l], J>0$ with 
$\chi$ the dimensionless spin-phonon coupling.

The equations of motion for the magnetic and elastic subsystems,
\begin{equation}
\frac{d}{dt}{\bf S}_n
=-{\bf S}_n\times\frac{\partial H}{\partial{\bf S}_n},~~~
M\frac{d^2}{dt^2} Q_n=-\frac{\partial H}{\partial Q_n},
\label{eqm}
\end{equation}
written in dimensionless form are, 
\begin{eqnarray}
\dot{\bf e}_n=c_t[(1-\chi\rho_{n-1})
{\bf e}_{n-1}\times{\bf e}_{n}
-(1-\chi\rho_n){\bf e}_{n}\times{\bf e}_{n+1}],
\label{eqmds} \\
\ddot{u}_n = c_e\chi(f_{n-1}-f_n)+V'(\rho_{n})-V'(\rho_{n-1})-U'(u_n),~~
\label{eqmdu}
\end{eqnarray}
where the dot denotes differentiation with respect to the 
dimensionless time $\tau=v_0t/l$, $\rho_n=u_{n+1}-u_n$, 
$u_n=Q_n/l$ is the dimensionless displacement and 
$f_n=({\bf e}_n\cdot{\bf e}_{n+1})$.
Here we have introduced, the coefficient $c_t=JSl/v_0$ as the ratio of 
characteristic elastic ($l/v_0$) and magnetic ($1/JS$) times and  
$c_e=JS^2/Mv_0^2$ as the ratio of characteristic 
magnetic ($JS^2$) and elastic ($Mv_0^2$) energies. In the following, 
we take $JS^2$ as the unit of energy.

We are going to consider three types of inter-cite potentials, 
(i) the harmonic potential $V(\rho)=\rho^2/2$, 
(ii) the FPU potential $V(\rho)=\rho^2/2+\rho^4/4$,
(iii) the Toda potential $V(\rho)=\exp(-\rho)+\rho-1$
and two type of on-cite substrates,
(iv) the harmonic $U(u)=k u^2/2$ 
and (v) the anharmonic sine-Gordon, $U(u)=1-\cos(2\pi u)$.

The framework for studying the thermal conductivity is the 
linear response Green-Kubo formalism where the thermal conductivity 
$\kappa$ at 
temperature $T$ is given by the energy current $J$ autocorrelation,
$C(\tau)=\lim\limits_{N\rightarrow\infty}C_N(\tau)$, 
$C_N(\tau)=\langle J(\tau)J(0)\rangle/NT^2$, 
\begin{equation}
\kappa =\lim_{\tau\rightarrow\infty}\int_{0}^\tau C(s)ds.
\label{kappa}
\end{equation}
$N$ is the number of particles in a chain with periodic boundary
conditions, $J(\tau)=\sum_{n=1}^N j_n(\tau)$ is the total energy current,  
with the local one $j_n$ obtained from the continuity equation for the 
energy; $j_n=\{g_{n-1}+g_n+(\dot{u}_{n-1}+\dot{u}_n)
[c_e\chi({\bf e}_{n-1}\cdot{\bf e}_n)
-V'(\rho_n)]\}/2$ and 
$g_n=c_tc_e(1-\chi\rho_{n-1})(1-\chi\rho_n)
({\bf e}_{n-1}\times{\bf e}_{n}\cdot{\bf e}_{n+1})$.
For $\chi=0$ we recover the usual expressions for the isolate spin and lattice 
energy current.
The averaging $\langle\cdot\rangle$ is performed over thermalized
states of the chain. 

To thermalize the system we use Langevin dynamics 
with the a white Gaussina noise thermostat coupled either to the lattice 
displacements only or to the spin and lattice displacements\cite{stz}. 
We verified that both approaches give practically the same results for 
thermodynamic (energy, specific heat) as well as dynamic quantities. 
Following thermalization, the integration of equations of motion 
(\ref{eqmds},\ref{eqmdu})
gives $C(\tau)$ up to times $\tau\sim 1000$ for $N=4000$ spin/atoms in a 
chain with periodic boundary conditions -- large enough to practically 
eliminate finite size effects. As $C(\tau)$ depends significantly on the 
concrete realization of the thermalized chain, we perform an average 
over about $10^4-10^5$ independent initial realizations to improve 
statistics. In all simulations we consider the parameters 
$c_t=1$ and $c_e=10$ that facilitate the convergence of the simulations but 
are also fairly realistic as the magnetic energy scale is often larger than 
the phonon scale.

Let's at first discuss the simplest case of a lattice with 
only harmonic inter-site interaction $V(\rho)=\rho^2/2$ 
coupled to spins. No substrate is present ($U(u)=0$) and thus the phonons 
are acoustic with linear low energy dispersion. 
As there is a zero mode with infinite lifetime due to the translational 
symmetry we take care to 
use thermalized states with zero center of mass motion 
$\sum_{n=1}^N \dot u_n(\tau)=0$.
\begin{figure}[tb]
\includegraphics[angle=0, width=.9\linewidth]{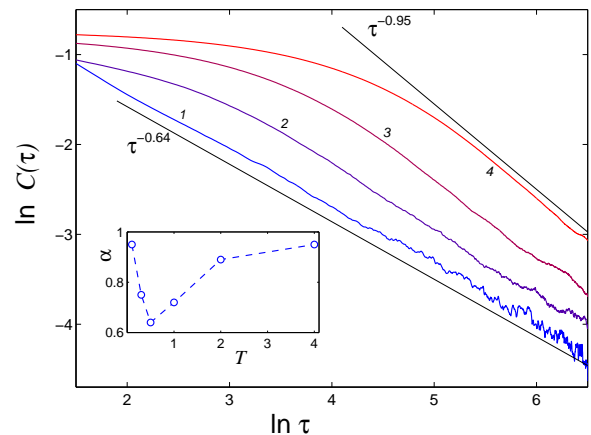}
\caption{
The power decay of the autocorrelation function $C(\tau) \sim \tau{-\alpha}$
in the spin-phonon chain with harmonic 
inter-cite potential (no substrate, $U(u)=0$) for $\chi=0.2$ and different 
temperatures: $T=$0.5, 1, 2, 4 (curves 1, 2, 3, 4). 
In the inset, temperature dependence of power $\alpha$.}
\label{fig1}
\end{figure}

If $\chi$ was zero, then the thermal conductivity would have two independent 
contributions. The first, 
from the phonon system that is diverging (purely ballistic) as the lattice is 
harmonic and the energy current is a constant of motion. 
The second, from the spins, 
that is finite (diffusive) with the spin energy current correlations decaying 
exponentially in time\cite{stz}. Thus $C(\tau)$ would decay exponentially fast 
to a constant (due to the phonon part) and the total thermal conductivity 
would of course be divergent.

In the spin-phonon coupled system ($\chi\ne 0$), 
as is shown in Fig. \ref{fig1}, the simulations indicate that  
the decay of the total energy current autocorrelations becomes 
power law $C(\tau)\sim \tau^{-\alpha}$ 
with $\alpha < 1$. Thus the thermal conductivity $\kappa$  of the 
coupled system -- diffusive spin plus ballistic phonon 
transport -- still diverges.
In the inset, the temperature dependence of $\alpha$ is shown.
It is nonmonotonic, with a minimum at $T\approx 0.5$, and $\alpha\rightarrow 
1$ as $T\rightarrow 0,\infty$, implying a logarithmically diverging 
conductivity at these limits.

\begin{figure}[tb]
\includegraphics[angle=0, width=0.9\linewidth]{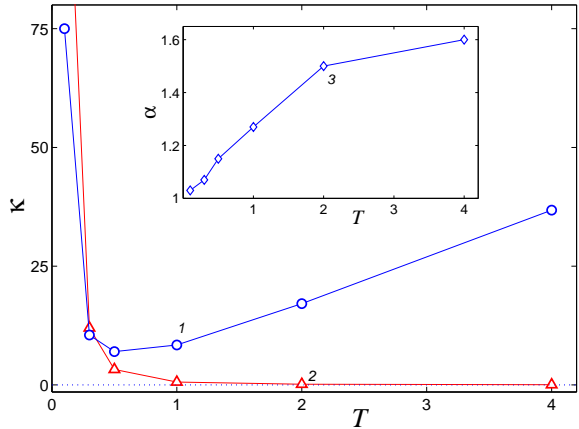}
\caption{Thermal conductivity $\kappa$ as a function of 
temperature for spin system coupled to harmonic inter-site and 
substrate potential (curve 1). 
For reference, thermal conductivity $\kappa_{sp}$ of uncoupled Heisenberg 
spin model (curve 2).
In the inset, temperature dependence of $\alpha$ in 
$C(\tau)\sim \tau^{-\alpha}$ - straight lines are aid to the eye.}
\label{fig2}
\end{figure}

In order to analyze mechanisms for breaking the ballistic behavior, 
we will now study phonon systems with substrate or with 
anharmonic inter-site interaction.
First, we consider the same harmonic inter-site potential but now in 
the presence of a substrate harmonic potential $U(u)=ku^2/2$ 
with $k=0.25$. Thus the phonon system remains harmonic and the corresponding 
phonon thermal transport ballistic. The same discussion as above applies 
for the case $\chi=0$.
Now however, as the simulations indicate, again a power law decay of the 
energy current autocorrelations is found $C(\tau)\sim\tau^{-\alpha}$ 
but with a power $\alpha > 1$ which 
implies a finite total thermal conductivity.
The temperature dependence of the spin-phonon thermal conductivity is shown 
in Fig. \ref{fig2}, along with the thermal conductivity of the spin 
system $\kappa_{sp}$ alone. We note that, unlike the spin thermal conductivity 
that decreases basically as $1/T^2$ at high $T$ due to the bounded spin 
spectrum, the total one increases linearly with $T$ at high 
temperatures (note that the uncoupled phonon part diverges). 
Thus it seems that the presence of a substrate is necessary for obtaining 
a finite $\kappa$ in the spin-phonon coupled system. 
The temperature dependence of $\alpha$ is shown in the 
inset indicating marginal, logarithmic convergence of the energy current 
autocorrelations 
as $T\rightarrow 0$ and a stronger one -- albeit still power law -- at high $T$.
At low temperatures, although the phonon contribution is divergent, 
the total thermal conductivity is lowered compared to the spin 
part. Thus the phonons basically act only as scatterers to the 
spin excitations.
Finally, note that as $c_e=10$, the ratio of magnetic to lattice energies, 
the phonon system is highly excited at $T/J\sim 1$ ($T/Mv_0^2\sim 10$).

\begin{figure}[t]
\includegraphics[angle=0, width=.9\linewidth]{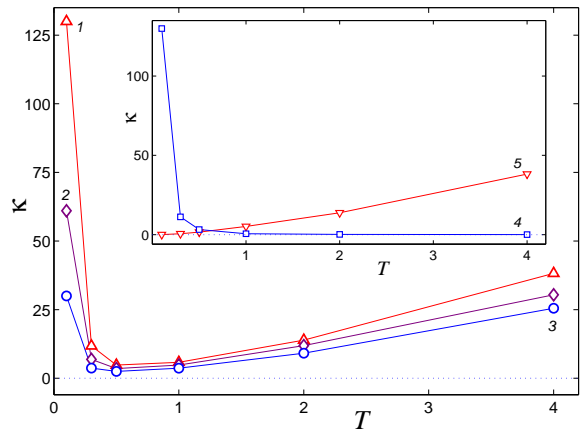}
\caption{Thermal conductivity $\kappa$ as a function of temperature 
for harmonic inter-site and anharmonic (periodic) substrate potentials for 
coupling $\chi=$0, 0.1, 0.2 (curve 1, 2, 3). The case $\chi=0$ corresponds to 
the sum of uncoupled spin $\kappa_{sp}$ and phonon $\kappa_{ph}$ contributions. 
In the inset, temperature dependence of individual 
$\kappa_{sp}$ and $\kappa_{ph}$ (curve 4 and 5).} 
\label{fig3}
\end{figure}

The above results were found using the Green-Kubo approach. 
Direct simulation of heat transport in a finite chain of length $N$, 
with ends in contact to heat reservoirs at different temperatures, gives 
good agreement between the two methods. 
In particular, for the harmonic chain without substrate the heat conductivity
coefficient diverges as $\kappa(N)\sim N^{0.34}$ for
$N\rightarrow\infty$ and $T=1$. For the chain with harmonic substrate
($k=0.25$) a finite limit exists, $\kappa=8.57$, compared to the value 
$\kappa=8.4$ obtained from (\ref{kappa}). 

To study the coupling of the diffusive spin system with a diffusive 
phonon one, we consider the on-site sine-Gordon potential 
$U(u)=1-\cos(2\pi u)$ with the same harmonic inter-site one 
$V(\rho)=\rho^2/2$. This anharmonic phonon system is 
known\cite{hlz,tbsz} to show normal (diffusive) thermal conductivity.
Thus for the uncoupled system $\chi=0$, the autocorrelations of 
each component decay exponentially with time and the thermal conductivities 
simply add. For $\chi\ne 0$, 
now the simulations show that the energy current autocorrelation 
of the spin-phonon coupled system (as the individual ones)  
decays exponentially $C(\tau)\sim e^{-\gamma\tau}$, $\gamma>0$, and the thermal conductivity 
is finite. In the inset of Fig. \ref{fig3} we show the temperature 
dependence of the 
individual ones for $\chi=0$ - $\kappa_{sp}$ for the spin, $\kappa_{ph}$ for 
the phonon subsystem. As we see $\kappa_{ph}$ increases 
monotonically with $T$ in the temperature region we are considering, while 
$\kappa_{sp}$ decreases roughly as $1/T^2$. Actually we expect that at lower 
temperatures $\kappa_{ph}$ increases again as the phonon system becomes 
asymptotically harmonic\cite{savin}. 
The thermal conductivity $\kappa$ of the coupled system, Fig. \ref{fig3},  
goes through a minimum at $T\approx 0.5$. 
It is interesting to note that 
the sum of individual component, $\kappa$ for $\chi=0$, is significantly larger 
than the actual $\kappa$ for $\chi\ne 0$ indicating the importance of 
spin-phonon scattering in limiting the thermal conductivity. The change 
$\delta\kappa/\kappa$ is 
largest, about $50-80\%$, at low temperatures and decreases saturating 
to $20-30\%$ at high temperatures.
\begin{figure}[t]
\includegraphics[angle=0, width=.85\linewidth]{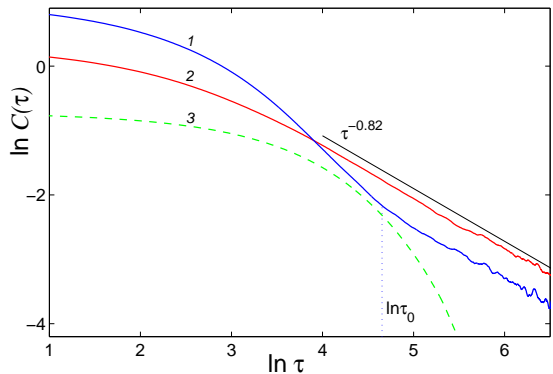}
\caption{
The power decay of the autocorrelation function $C(\tau)$ 
in the spin-phonon chain with FPU inter-cite potential  
(no substrate, $U(u)=0$) for $T=0.2$ and $\chi=0.2$ (curve 1).
For comparison, $C(\tau)$ for the isolated ($\chi=0$) 
FPU and spin chains (curves 2 and 3) also are given. 
}
\label{fig4}
\end{figure}

Finally, we will consider anharmonic phonon systems without substrate.
A classical example is the FPU chain\cite{livi} that has been extensively 
investigated and shows infinite conductivity. The autocorrelation
function $C(\tau)$ tends to zero as a power law $\tau^{-\alpha}$ 
with $\alpha<1$ that depends on temperature $T$. 
At high temperature $T=10$, $\alpha=0.63$ \cite{llp}, while for 
$T=0.2$, $\alpha=0.82$ (see Fig. \ref{fig4}). 
The spin-phonon system turns out to also show ballistic behavior.
As shown in Fig. \ref{fig4}, a difference to the FPU case is that here 
exist two time scales, $\tau<\tau_0$ where $C(\tau)$ decreases 
exponentially fast, turning to a power law for $\tau>\tau_0$.
For $T=0.2$, the switch time $\tau_0\approx 90$ and increases as the 
temperature tends to zero, a result of the dominating spin transport.
Note here, that the phonon FPU chain with
harmonic substrate has finite conductivity\cite{per} (power law decay 
of correlations with $\alpha >1$) and of course the corresponding 
spin-phonon system has too.

Another classic anharmonic phonon system is the Toda chain, an integrable 
system with ballistic transport\cite{xz}.
Our simulations (not shown) conclude that the spin-Toda chain has 
finite thermal conductivity with exponentially decaying correlations.
The temperature dependence of $\kappa$ is qualitatively similar to the one 
shown in Fig. \ref{fig3}. As the Toda chain with substrate potential 
(a nonintegrable system) becomes diffusive, we can speculate that the spin 
system acts as a substrate that breaks the integrability. 
Other spin-phonon examples with finite conductivity are the anharmonic 
Morse and Lenard-Jones potentials.

One surprizing outcome for the composite spin-phonon  system
with a  linear vibrational chain is that while
conductivity is infinite when the phonon chain is purely accoustic, it
becomes finite when there is a phonon gap.  Numerics shows that
in the high temperature regime thermalized spins act through the
term $c_e \chi ( f_{n-1} - f_n )$ of Eq. (\ref{eqmdu}) 
as dynamic scatterers for phonons with a power spectrum similar but not 
identical to that of white noise.
Specifically, we find that this effective additive noise term has
zero mean and is correlated with  $\tau_c = 0.5$.  In this high-temperature
regime one may use the analysis of mode coupling theory relating the energy 
autocorrelation to the phonon group velocity $v_g ( k ) = d w (k) / d k$ and
the effective local damping $\gamma ( k)$ as 
$C(\tau ) \sim \int_{- \pi }^{\pi} dk v^2 (k) 
e^{- \gamma (k) \tau }$\cite{livi}.
Algebraic dependence of damping on the wavevector $k$, viz.
$\gamma (k) \sim k^\delta$  results in algebraic decay in the
purely accoustic chain  as
$C (\tau ) \sim  \tau^{-1/\delta }$\cite{livi}, while
the presence of the gap results in dominant group
velocity contribution at some characteristic wavevector $k  \simeq  k_0$
leading asymptotically to an exponential decay
as $C (\tau ) \sim e^{- \gamma (k_0 ) \tau }$.  
This heuristic picture is compatible with the numerical evidence presented 
above.

In  conclusion we find that the thermal conductivity of
a  classical Heisenberg spin chain coupled
to linear or nonlinear phonon chains with a substrate 
is finite. The same holds true
when the spin system is coupled to anharmonic chains such as the
integrable  Toda lattice.  However, when the phonon chains is either
purely linear accoustic, or weakly nonlinear accoustic such as the 
FPU lattice  conductivity becomes anomalous. 
Further theoretical work is needed for the understanding of these 
numerical observations.

This work was supported by the E.U. grant MIRG-CT-2004-510543.
One of the us (A.V.S.) also thanks the Russian Foundation of
Basic Research (awards 04-02-17306) for partial financial support.


\begin{references}
\bibitem{hess} C. Hess, C. Baumann, U. Ammerahl, B. B\"uchner,
F. Heidrich-Meisner, W. Brenig, and A. Revcolevschi, Phys. Rev. B{\bf 64}, 
184305 (2001).
\bibitem{ott} 
A.V. Sologubenko and H.R. Ott, in {\it Strong Interactions in Low Dimensions}, 
Kluwer Academic Publishers , Dordrecht (2004);

\bibitem{review} X. Zotos, P. Prelov\v sek, {\it ibid}; arXiv:cond-mat/0304630.

\bibitem{shimshoni} E. Shimshoni, N. Andrei, and A. Rosch, 
Phys. Rev. B{\bf 68}, 104401 (2003).
\bibitem{chernyshev} A. V. Rozhkov and A. L. Chernyshev, 
Phys. Rev. Lett. {\bf 94}, 087201 (2005).
\bibitem{unpublished} P. Prelov\v sek, X. Zotos, unpublished.

\bibitem{livi} S. Lepri, R. Livi, A. Politi, Phys. Rep. {\bf 377}, 1 (2003).
\bibitem{muller} G. M\"uller, Phys. Rev. Lett. {\bf 60}, 2785 (1988).
\bibitem{stz} A. Savin, G. Tsironis and X. Zotos, 
Phys. Rev. B{\bf 72}, 140402 (2005).

\bibitem{hlz} B. Hu, B. Li and H. Zhao, Phys. Rev. E{\bf 57}, 2992 (1998).
\bibitem{tbsz} G.P. Tsironis, A.R. Bishop, A.V. Savin and A.V. Zolotaryuk, 
Phys. Rev. E{\bf 60}, 6610 (1999).

\bibitem{savin} A.V. Savin and O.V. Gendelman,
Phys. Rev. E{\bf 67}, 041205 (2003). 

\bibitem{llp} S. Lepri, R. Livi, and A. Politi, 
Europhys. Lett. {\bf 43} 271, (1998).
\bibitem{per} A. Pereverzev, Phys. Rev. E{\bf 68}, 056124 (2003).                     
\bibitem{xz} 
X. Zotos, J. Low Temp. Phys. {\bf 126}, 1185 (2002).

\end{references}
\end{document}